\documentclass[aps,prb,twocolumn,superscriptaddress]{revtex4-1}

\usepackage{graphicx}
\usepackage{color}

\begin{document}


\title{Temperature-dependent local structure of superconducting BaPd$_2$As$_2$ and SrPd$_2$As$_2$
}


\author{K. Terashima}
\email[]{k-terashima@cc.okayama-u.ac.jp}
\affiliation{Research Institute for Interdisciplinary Science, Okayama University, Okayama, 700-8530, Japan}

\author{E. Paris}
\altaffiliation{Research Department Synchrotron Radiation and Nanotechnology, Paul Scherrer Institut, CH-5232 Villigen PSI, Switzerland}
\affiliation{Dipartimento di Fisica, Universit\'{a} di Roma ``La Sapienza'' - P. le Aldo Moro 2, 00185, Roma, Italy}

\author{L. Simonelli}
\affiliation{CELLS - ALBA Synchrotron Radiation Facility, Carrer de la Llum 2-26, 08290, Cerdanyola del Valles, Barcelona, Spain}

\author{E. Salas-Colera}
\affiliation{Instituto de Ciencia de Materiales de Madrid, ICMM-CSIC, Sor Juana In\'es de la Cruz 3, 28049 Madrid, Spain}
\affiliation{Spanish CRG BM25 Spline, ESRF - The European Synchrotron, 71 avenue des Martyrs, 38043 Grenoble, France}

\author{A. Puri}
\affiliation{CRG-LISA, ESRF, 71 avenue des Martyrs, 38000 Grenoble, France}

\author{T. Wakita}
\affiliation{Research Institute for Interdisciplinary Science, Okayama University, Okayama, 700-8530, Japan}

\author{Y. Yamada}
\affiliation{Research Institute for Interdisciplinary Science, Okayama University, Okayama, 700-8530, Japan}

\author{S. Nakano}
\affiliation{Research Institute for Interdisciplinary Science, Okayama University, Okayama, 700-8530, Japan}

\author{H. Idei}
\affiliation{Research Institute for Interdisciplinary Science, Okayama University, Okayama, 700-8530, Japan}

\author{K. Kudo}
\affiliation{Research Institute for Interdisciplinary Science, Okayama University, Okayama, 700-8530, Japan}

\author{M. Nohara}
\affiliation{Research Institute for Interdisciplinary Science, Okayama University, Okayama, 700-8530, Japan}

\author{Y. Muraoka}
\affiliation{Research Institute for Interdisciplinary Science, Okayama University, Okayama, 700-8530, Japan}

\author{T. Mizokawa}
\affiliation{Department of Applied Physics, Waseda University, Tokyo, 169-8555, Japan}

\author{T. Yokoya}
\affiliation{Research Institute for Interdisciplinary Science, Okayama University, Okayama, 700-8530, Japan}

\author{N. L. Saini}
\affiliation{Dipartimento di Fisica, Universit\'{a} di Roma ``La Sapienza'' - P. le Aldo Moro 2, 00185, Roma, Italy}


\date{\today}

\begin{abstract}
The local structures of 122-type paradium arsenides, namely BaPd$_2$As$_2$ and SrPd$_2$As$_2$, are examined by As K-edge extended x-ray absorption fine structure measurements to find a possible correlation between the variation of their superconducting transition temperature and the local structure. The local atomic distances are found to be consistent with average distances measured by diffraction techniques. The temperature dependence of mean square relative displacements reveal that, while  BaPd$_2$As$_2$ is characterized by a local As-Pd soft mode, albeit with larger atomic disorder, SrPd$_2$As$_2$ shows anomalous As-Pd correlations with a kink at $\sim$160 K due to hardening by raising temperature. We have discussed implications of these results and possible mechanism of differing superconducting transition temperature in relation with the structural instability.
\end{abstract}

\pacs{pacs}

\maketitle
\section{INTRODUCTION}
Superconductivity in condensed matter generally appears or gets enhanced by a concomitant  suppression of the other orders, recognized as the interplay between different degrees of freedoms \cite{DagottoScience, ScalapinoRMP}.  In such cases, the knowledge of relation between crystal structure and superconductivity is of prime importance since this relationship largely affects other electronic degrees of freedoms. Iron-based superconductors are good example in which interplay between structural transition and superconductivity exists and the structural parameters have direct interplay with spin, charge and orbital-fluctuations, important for the superconductivity \cite{KontaniPRL}. Incidentally, such an intriguing interplay is strictly related to the As-chemistry, therefore interaction between  arsenic 4{\it p} and iron 3{\it d} electrons and bonding of As atoms \cite{NoharaAs}. The so-called ``122'' compounds \cite{Rotter, Sefat} forming a body-centered tetragonal ThCr$_2$Si$_2$-type structure (space group I4/mmm) at room temperature have been main systems to study these correlations \cite{Johnston, Si}. In general, the 122-type structure exhibits a structural phase transition from the tetragonal structure to an orhorhombic on cooling \cite{Huang}.  However, some compounds such as CaFe$_2$As$_2$ show other type of structural transition, namely a collapsed tetragonal (CT) phase transition under small external pressure, characterized by a discontinuous change in the material's lattice parameters and volume \cite{Goldman, Kreyssig}. The CT phase is nonmagnetic and lacks magnetic fluctuations with suppressed superconductivity \cite{Kreyssig, Goldman2}.  The driving mechanism of the CT phase has been attributed to the strong interlayer interaction through As-As covalent bonding \cite{Goldman2, Tsubota}.
	
As a matter of fact, superconductivity also appears in non-magnetic CT phase of 122 materials when Fe atoms are completely substituted by other metals, although the superconducting transition temperature ({\it T}$_c$) remains relatively low \cite{Bauer, Berry, Hirai1, Ronning, Hirai2, Kudo1, Anand, Guo, Wang, Kudo2}.  Recently, an interesting relationship between the Debye temperature and {\it T}$_c$ has been found in Pd-based 122-type of CT arsenide \cite{Kudo2}.  The compound with lower Debye temperature tend to show higher {\it T}$_c$.  For instance, BaPd$_2$As$_2$ shows superconductivity below $\sim$3.5 K with the Debye temperature being $\sim$140 K while {\it T}$_c$ of SrPd$_2$As$_2$ is $\sim$1 K with the Debye temperature of $\sim$300 K.  The estimated specific heat $\gamma$ values of these compounds are very close each other \cite{Kudo2}, suggesting that density of states of two compounds can be comparable. Hence it has been discussed that the presence of soft phonon may enhance electron-phonon interaction ($\lambda$) in the phonon-mediated superconductivity. In this context it should be important to know which local lattice mode may be important to drive this electron-phonon interaction in these materials.
	
In this work, we report a comparative temperature dependent study of the local structure of BaPd$_2$As$_2$ and SrPd$_2$As$_2$ by means of As K-edge extended x-ray absorption fine structure (EXAFS).  We have found a marked difference in the temperature dependence of mean square relative displacements (MSRD) of As-Pd bond in these two compounds. While the general trend of Einstein temperature ($\Theta_E$) in the two systems is consistent with that of the Debye temperature, we find that the MSRD of As-Pd bond in BaPd$_2$As$_2$ can be described by a single Einstein temperature, whereas that in SrPd$_2$As$_2$ shows crossover like behavior $\sim$160 K from lower to higher Einstein temperature.  We have also found that, although the {\it T}$_c$ of BaPd$_2$As$_2$ is more than three times higher than that of SrPd$_2$As$_2$, the static disorder in the latter is much lower than that in the former.  The results are consistent with structural instability induced phonon softening to trigger strong coupling superconductivity in BaPd$_2$As$_2$.

\section{METHODS}	
Polycrystalline samples of BaPd$_2$As$_2$ and SrPd$_2$As$_2$ were synthesized by heating starting materials in sealed quartz tubes.  It is known that BaPd$_2$As$_2$ has three polymorphs, namely, ThCr$_2$Si$_2$-type, CeMg$_2$Si$_2$-type, and alternately stacked CaBe$_2$Ge$_2$- and CeMg$_2$Si$_2$-type structures \cite{Mewis1, Mewis2}.  Among them, ThCr$_2$Si$_2$-type exhibits superconductivity \cite{Guo, Kudo2} while the CeMg$_2$Si$_2$-type does not \cite{Kudo2}.  In this study, the sample of BaPd$_2$As$_2$ with the ThCr$_2$Si$_2$-type structure was selectively synthesized by controlling the synthesis conditions \cite{Kudo2}, and a {\it T}$_c$ value of 3.85 K was evaluated through the magnetization measurement. On the other hand, the {\it T}$_c$ value of SrPd$_2$As$_2$ was not evaluated because of its low {\it T}$_c$ (= 0.92 K) \cite{Anand}.
	
\begin{figure}
\includegraphics[width=8.6cm]{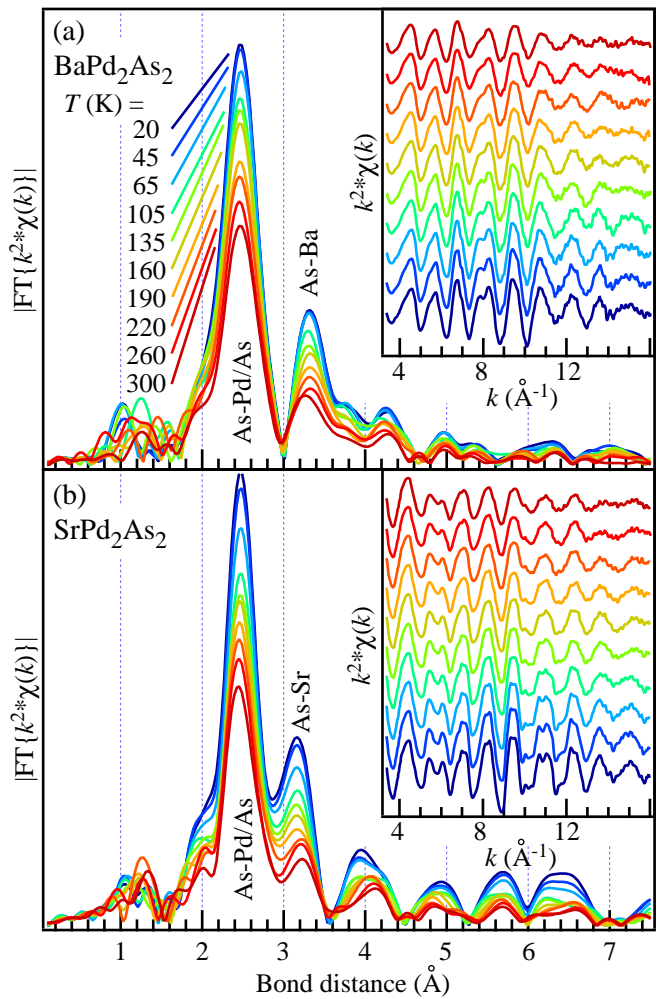}
\caption{(Color online) Temperature dependence of Fourier transform magnitudes of $k^2$-weighted As K-edge EXAFS on (a) BaPd$_2$As$_2$ and (b) SrPd$_2$As$_2$.  Inset shows the corresponding EXAFS oscillations.  The data are not corrected by the phase shifts.}.
\end{figure}

\begin{figure}
\includegraphics[width=8.6cm]{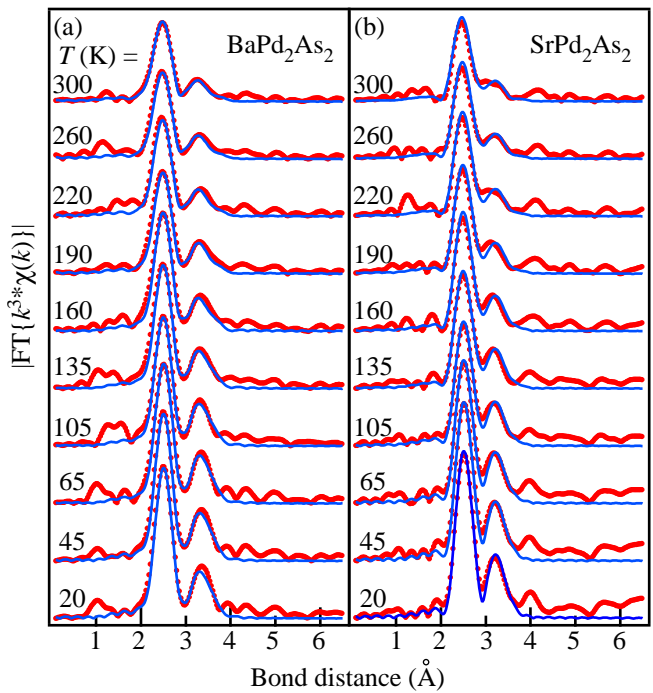}
\caption{(Color online) Fourier transform magnitudes of $k^3$-weighted As K-edge EXAFS on (a) BaPd$_2$As$_2$ and (b) SrPd$_2$As$_2$ after phase shift corrections.  Red dots represent experimental data and blue lines show the model fits.}
\end{figure}	
	
Arsenic K-edge (11868 eV) x-ray absorption measurements were performed at the Spline beamline \cite{Spline} of the European Synchrotron Radiation Facility, where the synchrotron radiation was monochromatized using a double crystal Si(111) monochromator.
The spectra were acquired sequentially on the two samples as a function of temperature in transmission mode using three ionization chambers mounted in series for simultaneous measurements on the sample and a reference. Finely powdered samples of BaPd$_2$As$_2$ and SrPd$_2$As$_2$ were mixed uniformly in an organic matrix and pressed into pellets of 13 mm diameter for obtaining the edge jumps to be about 1. Several absorption scans at each temperature (3-5 scans) were measured to ensure high signal to noise ratio and spectral reproducibility. The sample temperature during the measurements was controlled and monitored within an accuracy of $\pm$1 K. As K-edge x-ray absorption spectra were also measured on CaPd$_2$As$_2$ (see supplemental material \cite{supple}), a related compound, using the LISA beamline of the European Synchrotron Radiation Facility following the similar experimental approach.  The EXAFS oscillations were extracted from the absorption spectra using standard procedure based on spline fits \cite{Bunker}.
	
\section{RESULTS}

Figure 1 shows the temperature dependence of Fourier transform (FT) magnitudes of $k^2$-weighted EXAFS oscillations extracted from As K-edge x-ray absorption spectra measured on BaPd$_2$As$_2$ (a) and SrPd$_2$As$_2$ (b).  Insets show the corresponding EXAFS oscillations.  The data are not corrected by photoelectron phase shifts and hence represent raw experimental data. The EXAFS oscillations are visible up to high {\it k}-range ($\sim 16 \AA^{-1}$) even at 300 K. In both compounds, the first FT peak appears at $\sim$2.5 {\rm \AA} that corresponds to Pd-As and As-As bonds, and the second peak at 3.2-3.4 {\rm \AA} corresponds to As-Ba or As-Sr distance.  It is clear from the figure that the local bond distances in SrPd$_2$As$_2$ are shorter than those in BaPd$_2$As$_2$, which is in general agreement with those estimated from x-ray diffraction measurements \cite{Mewis2, Hofmann}.  The peaks appear $\geq$4 {\rm \AA} would originate from multiple scattering as well as distant atom contributions.  The FT peak amplitudes $\geq$4 {\rm \AA} tend to be higher in SrPd$_2$As$_2$ than in BaPd$_2$As$_2$, indicating smaller overall disorder in SrPd$_2$As$_2$.
	
To quantify the local lattice parameters, the measured EXAFS oscillations were modeled with three shells involving the nearest neighbor As-Pd, As-As, and As-Ba/Sr distances. The EXAFS modeling was carried out in the single scattering approximation using the standard EXAFS equation \cite{Bunker}. The EXCURVE 9.275 code \cite{EXCURVE} was used for the model fits in which calculated amplitude and phase factors were used. In the starting model the structural parameters deduced from x-ray diffraction measurements were used \cite{Mewis2, Hofmann}. In these model fits, the passive electrons reduction factor $S_0^2$ was set to 0.9, while the number of neighboring atoms $N_i$ were fixed to their nominal values (4 for As-Pd, 1 for As-As, 4 for As-Ba/Sr).  The photoelectron energy zero ({\it E}$_0$) was fixed after fit trials on several scans. The fitting {\it k}-range was 3.4-16 ${\rm \AA^{-1}}$. The number of independent data points, 2$\Delta k\Delta R/\pi$, for the present EXAFS model fits ($\Delta$ {\it R} = 2.0 ${\rm \AA}$) was $\sim$16 in which eventually six parameters were varied.  It should be mentioned that apart from sequential measurements and the same data treatments, the EXAFS model fit approach was same for the two systems for a direct comparison of the derived physical parameters. The uncertainties in the physical parameters were determined by creating correlation maps and by analyzing independent scans.  Figure 2 shows the temperature dependent FT of {\it k}$^3$-weighted As K-edge EXAFS of BaPd$_2$As$_2$ (a) and SrPd$_2$As$_2$ (b) after phase shift correction along with the three-shells model fits. The spectra are displayed as red dots with offset for each temperature, and the vertical scale of Figs. 2(a) and 2(b) are kept to be the same for a realistic comparison. 

\begin{figure}[t]
\includegraphics[width=8.6cm]{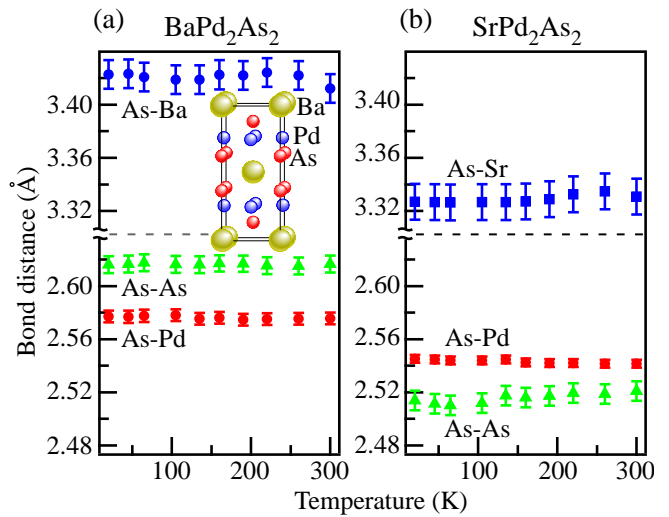}
\caption{(Color online) Experimentally determined local bond distances around As atom of (a) BaPd$_2$As$_2$ and (b) SrPd$_2$As$_2$.  Inset in (a) shows the crystal structure.}
\end{figure}

\begin{figure}[t]
\includegraphics[width=8.0cm]{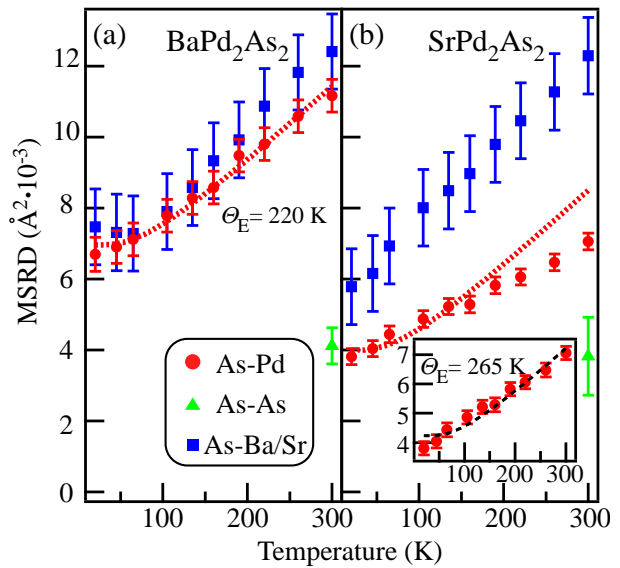}
\caption{(Color online) MSRDs of As-Pd (red circle) and As-As (green triangle), and As-Ba/Sr (blue square) on (a) BaPd$_2$As$_2$ and (b) SrPd$_2$As$_2$.  Red dotted lines denote results of the Einstein model fit for As-Pd bond.  In (b), the fit was performed in a temperature range of 20-160 K with the same Einstein-frequency with (a) (220 K).  Inset of (b) shows the fit result with Einstein-frequency as a free parameter.}
\end{figure}
		
Figure 3 shows the temperature dependence of bond distances from As atoms for BaPd$_2$As$_2$ (a) and SrPd$_2$As$_2$ (b).  In both compounds, all the bond distances at room temperature appears in an overall agreement with those reported by x-ray diffraction \cite{Mewis2, Hofmann}.  As we change the Alkaline earth metal from Ba to Sr, all bond distances tend to contract.  This is qualitatively reasonable, considering their ionic radius $\sim$1.61 {\rm \AA} for Ba and $\sim$1.44 {\rm \AA} for Sr in XII coordination \cite{Shannon}.  Among three obtained bonds, As-As distance showed the largest decrease of about 5 $\%$.  The As-As distance in BaPd$_2$As$_2$ ($\sim$2.6 {\rm \AA}) is comparable to those of collapsed phase of 122-type iron-arsenides.  As we substitute Ba with Sr, the lattice constant shrinks and As-As distance is further compressed.  
	
We show in Fig. 4 the temperature dependence of MSRD, where we have found interesting differences between two materials.  While MSRD of As-Pd bond in BaPd$_2$As$_2$ (Fig. 4(a)) monotonically increases by raising temperature, that in SrPd$_2$As$_2$ (Fig. 4(b)) shows a kink structure $\sim$160 K.  On the other hand, the MSRDs of As-Ba/Sr bonds in two compounds show similar behavior including their magnitude.  For As-As, we show the estimated values only for the highest temperature as a representative since we found that it has relatively large error partially due to small number of neighboring atoms with signal largely overlapped with As-Pd signal.  We have evaluated the Einstein temperatures, {\it $\Theta_E$}, of As-Pd bond using the correlated Einstein-model \cite{Sevillano}; $\sigma^2$ = $\sigma_0^2$ + $\sigma^2$({\it T}) where $\sigma_0^2$ is static configulational disorder and $\sigma^2$({\it T}) describes dynamic part.  The obtained {\it $\Theta_E$} value of As-Pd atomic pairs in BaPd$_2$As$_2$ was $\sim$220 K.  For SrPd$_2$As$_2$, first we tried to fit the temperature dependent MSRD of As-Pd using a single Einstein temperature (see, e.g., inset of Fig. 4(b)) and obtained a value of $\sim$265 K with $\sigma_0^2$$\sim$2.2$\times$10$^{-3}$ {\rm \AA}$^2$ , but the fitting was unsatisfactory especially at low-{\it T} side due to the presence of a kink structure in MSRD.  Instead, we have found that the data in the low temperature range (20-160 K) can be well described by the same Einstein temperature with that of BaPd$_2$As$_2$ as shown in Fig. 4(b).  As temperature raises more than $\sim$160 K, the MSRD of As-Pd bond in SrPd$_2$As$_2$ deviates from the expected behavior of {\it $\Theta_E$} = 220 K and it turns to a harder local mode.  Indeed, the estimated {\it $\Theta_E$} for {\it T}$>$160 K was $\sim$295 K. In addition, the evaluated $\sigma_0^2$ values for the As-Pd  correlations were found to be $\sim$4.5$\times$10$^{-3}$ {\rm \AA}$^2$ for BaPd$_2$As$_2$, $\sim$1.6$\times$10$^{-3}$ {\rm \AA}$^2$ for low-{\it T} part of SrPd$_2$As$_2$, and $\sim$2.9$\times$10$^{-3}$ {\rm \AA}$^2$ for high-{\it T} part of SrPd$_2$As$_2$.

\section{DISCUSSION}	
			
Here we discuss the possible implications of the present results.  Firstly the MSRDs of As-Ba/Sr bonds in two compounds turned out to be quite similar for the two systems with small differences in the configurational disorder, larger in BaPd$_2$As$_2$. Therefore, the local mode involving alkaline earth metals may not have a direct impact on superconductivity of this system.  On the other hand, we found two differences in MSRDs of As-Pd bonds in BaPd$_2$As$_2$ and SrPd$_2$As$_2$;  (i) the $\sigma_0^2$, describing static configurational disorder at low-{\it T} and (ii) the Einstein temperature in high-{\it T} region.  It seems that at the local scale, As-Pd correlations in SrPd$_2$As$_2$ goes through a lower {\it $\Theta_E$} to higher {\it $\Theta_E$}, i.e. a tendency from softer to harder local mode involving As-Pd correlations.  The latter appears consistent with the measured trend of the Debye temperature by the specific heat.  We stress that the static disorder ($\sigma_0^2$) in BaPd$_2$As$_2$ with {\it T}$_c$ = 3.85 K is more than twice of that in SrPd$_2$As$_2$ with {\it T}$_c$ $\sim$1 K.  Naively one expects that higher disorder should suppress the transition temperature, but apparently this is not the case.  Therefore, the configulational disorder may not be coming from any impurity but it should be associated with some kind of structural instability.   Incidentally, as mentioned earlier, BaPd$_2$As$_2$ is known to have polymorphs \cite{Mewis1, Mewis2, Kudo2}, while SrPd$_2$As$_2$ stabilizes in ThCr$_2$Si$_2$ structure.  The first principles calculations \cite{Shein} suggest that non-superconducting CeMg$_2$Si$_2$-type is more stable than superconducting ThCr$_2$Si$_2$-type structure in BaPd$_2$As$_2$.  The polymorphism is expected to induce larger configulational disorder and hence softer As-Pd bonds as described by lower Einstein temperature in BaPd$_2$As$_2$. On the other hand,  SrPd$_2$As$_2$  does not have such polymorphism but it goes through a local configurational transition at $\sim$ 160 K. Therefore, the superconductivity in BaPd$_2$As$_2$ is likely to be due to structural instability induced softening, that may lead to a strong electron-phonon coupling. It should be mentioned that the isostructural CaPd$_2$As$_2$ is superconducting with a $T_c$ of 1.27 K and a Debye temperature of $\sim$276 K\cite{Anand}, and can be considered intermediate between BaPd$_2$As$_2$ and SrPd$_2$As$_2$ systems. The Einstein temperature of As-Pd bond in CaPd$_2$As$_2$, determined by temperature dependent As K-edge EXAFS, is found to be $\sim$250 K (see the supplemental material \cite{supple}), consistent with the avobe arguments. Furthermore, the substitution of phosphorus for arsenic atoms suppresses the structural transition and raises {\it T}$_c$ in a sister compound BaNi$_2$As$_{2-x}$P$_x$  \cite{Kudo1}. It would be interesting to examine if similar mechanism is valid for the increased {\it T}$_c$ also in BaNi$_2$As$_{2-x}$P$_x$, which has to be clarified in future work.

\section{CONCLUSION}

In summary, we have studied the local structure of 122-type palladium-arsenides with different {\it T}$_c$s.  A sistematic temperature dependent study has permitted us to underline marked differences in the mean squrare relative displacement of As-Pd bonds for both static disorder as well in bond stiffness determined by the the related Eienstein temperatures.  BaPd$_2$As$_2$ has a relatively large static configurational disorder with a single Einstein temperature, while SrPd$_2$As$_2$ contains less static disorder with an anomalous change in the Einstein temperature, from softer to harder by raising temperature. We have argued that structural instability in BaPd$_2$As$_2$ is the likely reason for the existence of the soft phonon producing strong-coupling superconductivity.

\begin{acknowledgments}
	We thank ESRF staff for support in the EXAFS data collection.  K. T. and T. W. would like to acknowledge the hospitality at the Sapienza University of Rome.  This research was partially supported by the Program for Promoting the Enhancement of Research University from MEXT, the Program for Advancing Strategic International Networks to Accelerate the Circulation of Talented Researchers from JSPS (R2705), and JSPS KAKENHI (15H05886, 15K21732, 16K05451).  This work is a part of the executive protocol of the general agreement for cooperation between the Sapienza University of Rome and Okayama University, Japan.
\end{acknowledgments}


\clearpage
\widetext
\begin{center}
\textbf{\large Supplemental Information}
\end{center}

\setcounter{figure}{0}
\setcounter{section}{0}
\setcounter{page}{1}
\renewcommand{\thefigure}{S\arabic{figure}}
\renewcommand{\thetable}{S\arabic{table}}

\section{The local structure of CaPd$_2$As$_2$ measured by temperature dependent As K-edge EXAFS}
	CaPd$_2$As$_2$ is a related compound of (Ba/Sr)Pd$_2$As$_2$ with ThCr$_2$Si$_2$ structure.  $T_c$ of CaPd$_2$As$_2$ has been reported to be 1.27 K and estimated Debye temperature is 276 K, thus the material can be regarded to be intermediate to BaPd$_2$As$_2$ and SrPd$_2$As$_2$ \cite{KudoJPSJ}.  Figure S1(a) shows Fourier transform magnitudes of $k^3$-weighted EXAFS oscillations extracted from As K-edge x-ray absorption spectrum (inset) measured on CaPd$_2$As$_2$ at $T$ = 20 K, after phase shift corrections.  The solid line in the figure is the model fit, which was perfomed in the same way as in the main text, except for the fact that the third shell here is As-Ca instead of As-Ba/Sr.  The estimated bond distances and Einstein-frequency of As-Pd bond are tabulated in Table S1, along with $T_c$s and Debye temperatures taken from literature \cite{KudoJPSJ, Anand}.  The derived local bond distances are consistent with the values in the averaged structure reported earlier \cite{Mewis}.  In Fig. S1(b), the red dashed line show the fit result of MSRD in a temperature range of 20-160 K with the same Einstein-frequency with that of BaPd$_2$As$_2$ (220 K), while inset shows the fit result using all the temperature range with Einstein-frequency as a free parameter.

\begin{figure}[h]
\includegraphics[width=10 cm]{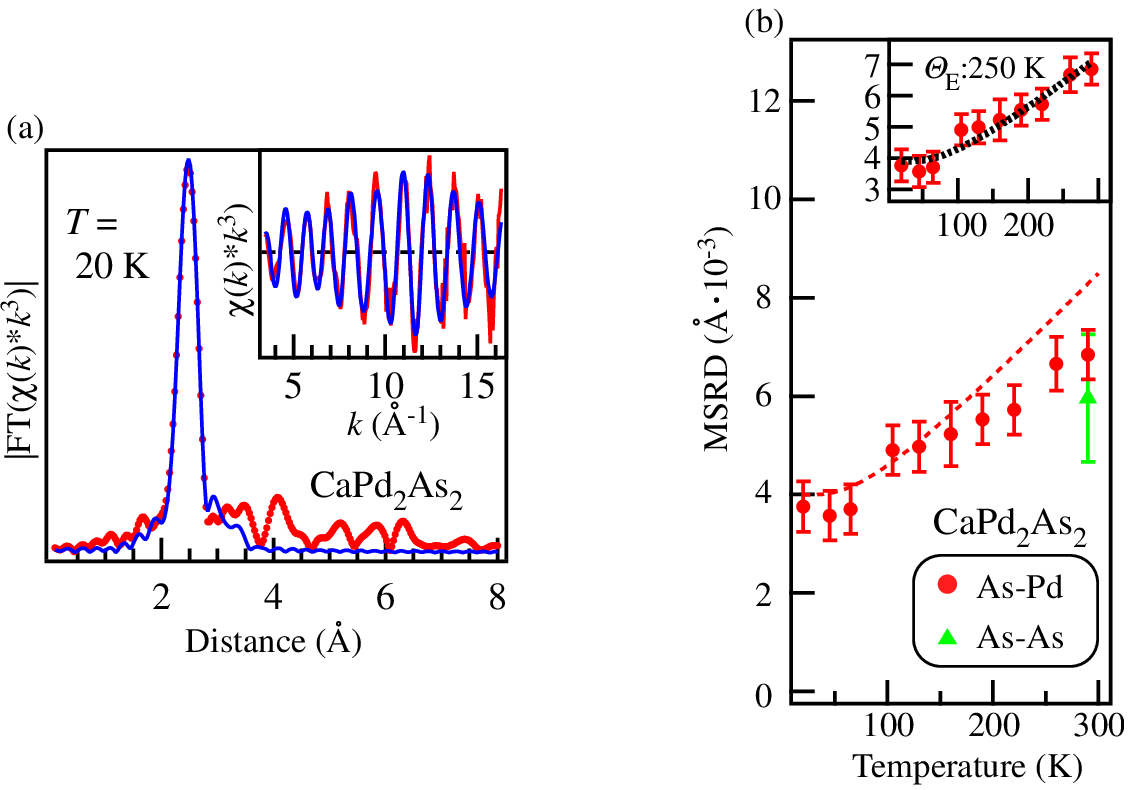}%
\caption{ (a) Fourier transform magnitudes of $k^3$-weighted As K-edge EXAFS on CaPd$_2$As$_2$ after phase shift corrections. Inset shows the $k^3$-weighted EXAFS oscillations of CaPd$_2$As$_2$ at $T$ = 20 K.  Red dots represent experimental data and blue lines show the model fits.  (b) Obtained MSRD of As-Pd bond as a function of temperature.  Dotted lines denote results of the Einstein model fit for As-Pd bond.  The fit was performed in a temperature range of 20-160 K with the same Einstein-frequency with BaPd$_2$As$_2$ (220 K).  Inset of (b) shows the fit result with Einstein-frequency as a free parameter.
}
\end{figure}

\begin{table}[h]
\caption{Table of physical and structural parameters in (Ba, Ca, Sr)Pd$_2$As$_2$. $\Theta$$_{E}$ and $\sigma$$_0$$^2$ show the values estimated for As-Pd bond using the data in the same temperature range of 20-300 K.  Values of local bond distances were taken from $T$ = 20 K data.}
\begin{tabular}{cccccccc}\hline

 				& $T_c$ (K) 			& $\Theta$$_D$ (K)		& $\Theta$$_{E}$ (K)		& $\sigma$$_0$$^2$ ($\AA$$^2$)	& $R_{As-Pd}$ ($\AA$)	& $R_{As-Pd}$	($\AA$)	& $R_{As-Pd}$ ($\AA$)\\ \hline
BaPd$_2$As$_2$	& 3.5$^{[1]}$	& 144$^{[1]}$		& 220				& 4.5$\times$10$^{-3}$			& 2.58					& 2.62				& 3.42			\\
CaPd$_2$As$_2$	& 1.27$^{[2]}$	& 276$^{[2]}$		& 250				& 1.7$\times$10$^{-3}$			& 2.52					& 2.44				& 3.26			\\
SrPd$_2$As$_2$	&0.92$^{[2]}$	& 298$^{[2]}$		& 265				& 2.2$\times$10$^{-3}$			& 2.55					& 2.51				& 3.33			\\	\hline
 \end{tabular}
\end{table}

\end{document}